\documentclass[twoside,11pt]{article}
\usepackage{algorithm}
\usepackage[preprint]{jmlr2e}
\usepackage[T1]{fontenc}
\usepackage{lmodern,microtype,amsmath,mathtools,booktabs,array,enumitem,xcolor}
\usepackage[noend]{algpseudocode}
\hypersetup{colorlinks=true,linkcolor=blue!45!black,citecolor=blue!45!black,urlcolor=blue!45!black}
\newcommand{\OPT}{\operatorname{OPT}}
\newcommand{\E}{\mathbb E}
\newcommand{\eps}{\varepsilon}
\newcommand{\one}{\mathbf 1}
\newcommand{\aoff}{\alpha_{\mathrm{off}}}
\newcommand{\inner}[2]{\langle #1,#2\rangle}
\newcommand{\symdiff}{\triangle}
\makeatletter
\let\lemma\@undefined\let\endlemma\@undefined
\let\thelemma\@undefined\let\c@lemma\@undefined
\makeatother
\newtheorem{lemma}{Lemma}

\setlist{itemsep=4pt,topsep=5pt}
\renewenvironment{proof}{\par\noindent\textbf{Proof.}\ }{\hfill\BlackBox\par\medskip}
\ShortHeadings{}{}
\firstpageno{1}
\begin{document}
\title{Tight Bounds for Non-Monotone Submodular Maximization\\
under a Cardinality Constraint with Constant Recourse}
\author{\name Vaneet Aggarwal\\
        \addr Purdue University\\
        West Lafayette, IN 47907, USA}
\maketitle
\begin{abstract}
We study the maximization of a nonnegative submodular function under a
cardinality constraint when elements arrive one at a time and an
algorithm must report a feasible set after every arrival. The recourse
of an algorithm is the largest number of insertions and deletions that
it makes to the reported set at a single arrival. For every $\eps>0$,
we give an algorithm with recourse $O(\eps^{-2})$ whose reported set
has, at every time, expected value at least $1/4-\eps$ times the largest
value of a feasible set of arrived elements. The algorithm evaluates the
function polynomially often, and only on sets of arrived elements. We
also show that no algorithm with constant recourse achieves a ratio
above $1/4$, even if it knows the function and the arrival order in
advance and has unlimited computation, and that any fixed ratio above
$1/4$ requires recourse that grows as a positive power of the
cardinality budget. Thus $1/4$ is the optimal limiting ratio with
constant recourse. The previous best ratio with constant recourse was
$3-2\sqrt2\approx0.172$. Our algorithm assigns selection probabilities
of at most one half and maintains, as elements arrive, an inequality
bounding the value of every feasible set by a multiple of the expected
value of a random set drawn with these probabilities; the reported set
moves to a rounding of these probabilities through gradual
replacements. The upper bound uses a weighted directed graph whose
vertices arrive in groups of decreasing sizes.
\end{abstract}

\section{Introduction}\label{sec:intro}
In many applications a selected set must be useful after every arrival of
new data, and not only at the end of the stream. Examples are a summary
shown to users, a training subset, and a collection of cached items.
Replacing a selected item has a cost, so the number of changes made at
each arrival should stay bounded. We measure the value of a set by a
submodular function, which has diminishing returns: adding an element to
a set gains no more than adding it to a subset of that set. We do not
assume that the function is monotone. In a weighted directed graph, for
example, the total weight of the arcs leaving a set is submodular, and
adding the head of such an arc to the set removes the arc's weight. For
objectives of this kind, an algorithm may have to delete old elements as
well as insert new ones.

Let $V$ be the set of elements in arrival order, and let
$f:2^V\to\mathbb R_{\ge0}$ be submodular. After
$t$ arrivals, an algorithm reports a set $S_t$ of at most $k$ arrived
elements, and $\OPT_t$ is the largest value of such a set. The recourse
at time $t$ is the number $|S_t\symdiff S_{t-1}|$ of insertions and
deletions. We seek an algorithm with $\E f(S_t)\ge\alpha\OPT_t$ at every
time $t$ whose recourse at every arrival is at most a constant. The
constant may depend on an accuracy parameter, but not on $k$ or on the
number of arrivals. The algorithm accesses $f$ through value queries,
each of which returns $f(A)$ for a set $A$ of arrived elements chosen by
the algorithm.

Two earlier approaches show the tension between the ratio and the
recourse. Sample-Streaming of \citet{feldman18} achieves the ratio
$3-2\sqrt2\approx0.172$ under a cardinality constraint with at most two
changes per arrival. Running the offline algorithm of \citet{aggdu26} on
every prefix achieves $\aoff-\eps$, where
\[
 \aoff=\frac{54440750000}{117486474169}=0.463378873\ldots,
\]
but two consecutive outputs can differ in $2k$ elements. For monotone
objectives, \citet{fu26} showed that $2-\sqrt2$ is the best ratio
achievable with constant recourse and polynomially many queries. For
non-monotone objectives, adding recent arrivals to a previously prepared
set can lower its value, so their analysis does not carry over.

We show that $1/4$ is the optimal limiting ratio with constant recourse.
Theorem~\ref{thm:poly-quarter} gives an algorithm that achieves
$1/4-\eps$ with recourse $O(\eps^{-2})$ and polynomially many queries on
arrived elements. Theorem~\ref{thm:quarter-upper} shows that no
algorithm with constant recourse exceeds $1/4$, even if it knows the
function and the arrival order in advance and makes unlimited queries.
Corollary~\ref{cor:poly-recourse} shows that any fixed ratio above $1/4$
requires recourse that grows as a positive power of $k$.
Table~\ref{tab:summary} places these bounds next to those for the
setting without a recourse limit. The ratio $1/4$ is about $45.7\%$
larger than $3-2\sqrt2$, and the recourse constant grows as $\eps$
decreases.

Our algorithm (Section~\ref{sec:algorithm}) assigns every arrived
element a selection probability of at most one half; we call this vector
the fractional solution. At random times, called checkpoints, the
algorithm recomputes the fractional solution on all arrived elements
under a budget slightly below $k$, so that no feasible direction improves
it by much. This property gives an inequality that bounds the gain that
any feasible set can add to the fractional solution by a multiple of the
expected value of the fractional solution
(Lemma~\ref{lem:dual-certificate}). When the estimated marginal value
of a new element exceeds its error tolerance, the algorithm assigns it
probability one half. If the estimates are accurate, this raises the
expected value by at least as much as it raises the bound on the gain,
and the marginal values of the old elements can only decrease; the
estimation errors are accounted for separately.
The inequality therefore continues to give a ratio close to $1/4$ at
every later time, even if the optimum grows by a large factor
(Lemma~\ref{lem:snapshot}). The reported set follows a random rounding
of the fractional solution. A new rounding can differ from the reported
set in up to $2k$ elements, so the algorithm carries out the replacement
over several arrivals, as in the checkpoint method of
\citet{duetting25,fu26}, and the budget reserved below $k$ keeps the set
feasible during the replacement.

The upper bound (Section~\ref{sec:upper}) adapts a construction of
\citet[Appendix A]{buchbinder19} for online algorithms that cannot remove
accepted elements. In their instance, every arrival is a vertex of a
weighted directed graph. We replace each vertex by a group of elements
and make every group much smaller than the previous one. With bounded
recourse, all changes made after a group arrives alter only a small
fraction of that group, so the fraction selected from the group is
essentially fixed once the group has arrived. Arc weights grow from group
to group, so the newest group accounts for most of the value when it
completes. An averaging argument over the selected fractions then bounds
the ratio by $1/4$. The graph and the arrival order are fixed in advance,
so the bound holds for algorithms that know them.

\begin{table}[t]
\centering
\caption{Achievable ratios and upper bounds for non-monotone submodular
maximization under a cardinality constraint, with the source of each
entry below it. All ratios are in expectation at each fixed time, and
recourse counts both insertions and deletions. Without a limit on
recourse, the ratio is attained by applying Corollary~6.5 of
\citet{aggdu26} to each arrived prefix, and consecutive outputs can
differ in up to $2k$ elements; the upper bound is Lemma~8.5 of
\citet{agglu26}. With constant recourse, $3-2\sqrt2$ is
the previous ratio, attained by Sample-Streaming with at most two changes
per arrival. Since the algorithm of Theorem~\ref{thm:poly-quarter} runs in
the most restricted model and the upper bound of
Theorem~\ref{thm:quarter-upper} holds in the least restricted one, $1/4$
is the limiting ratio in every model between them.}\label{tab:summary}
\small
\renewcommand{\arraystretch}{1.15}
\newcommand{\entry}[2]{\begin{tabular}[t]{@{}c@{}}#1\\[-1pt]
 {\footnotesize #2}\end{tabular}}
\begin{tabular}{@{}>{\raggedright\arraybackslash}p{0.95in}
 >{\raggedright\arraybackslash}p{1.25in} c c@{}}
\toprule
Setting & Algorithm and queries & Achievable ratio & Upper bound\\
\midrule
No limit on recourse
 & Polynomial queries
 & \entry{$0.463378873\ldots-\eps$}{\citet{aggdu26}}
 & \entry{$0.470438681\ldots$}{\citet{agglu26}}\\
\midrule
Constant recourse
 & Polynomial queries on arrived elements
 & \begin{tabular}[t]{@{}c@{}}$3-2\sqrt2\approx0.172$\\[-1pt]
   {\footnotesize\citet{feldman18}}\\[3pt]
   $1/4-\eps$\\[-1pt]
   {\footnotesize Theorem~\ref{thm:poly-quarter}}\end{tabular}
 & \begin{tabular}[t]{@{}c@{}}$1/4$\\[-1pt]
   {\footnotesize Theorem~\ref{thm:quarter-upper}, even with}\\[-2pt]
   {\footnotesize full foresight and}\\[-2pt]
   {\footnotesize unrestricted queries}\end{tabular}\\
\bottomrule
\end{tabular}

\end{table}

\section{Related work}\label{sec:related}

\paragraph{Online algorithms with removals.}
\citet{buchbinder19} study online algorithms that may accept an element
only when it arrives and may later remove it. Their ratio under a
cardinality constraint is $56/627-\eps\approx0.0893-\eps$.
Sample-Streaming improves the ratio to $3-2\sqrt2$ for a matroid
constraint, and hence for a cardinality constraint
\citep[Theorem 1]{feldman18}. Under a cardinality constraint, each
exchange step of Sample-Streaming inserts at most one element and removes
at most one, and the algorithm does not use the stream length, so it
achieves $3-2\sqrt2$ at every time with recourse at most two. Both
algorithms store few elements. Our model lets an algorithm store all
arrived elements and reinsert removed ones, and our algorithm uses this
to recompute its fractional solution on all arrived elements.

\paragraph{Consistent maximization and knowledge of the future.}
\citet{duetting24,duetting25} study algorithms for monotone submodular
maximization that bound the changes to the reported set, and they
introduce a checkpoint method that replaces a prepared set gradually.
\citet{fu26} prove that $2-\sqrt2$ is the best ratio for monotone
objectives with polynomially many queries and constant recourse measured
by symmetric difference. With full knowledge of the future and unlimited
computation, \citet{balkanski26} obtain the larger threshold of about
$0.7942$ for monotone objectives. For non-monotone objectives, old
elements that remain in the reported set during a replacement can lower
the value of the new set. Our schedule therefore deletes old elements
before inserting new ones, and our analysis follows a fractional solution
that does not depend on the reported set. By
Theorem~\ref{thm:quarter-upper}, knowledge of the future does not raise
the threshold above $1/4$ for non-monotone objectives.

\paragraph{Offline ratios and amortized recourse.}
\citet{aggdu26} achieve $\aoff-\eps$ for non-monotone maximization under
a cardinality constraint with polynomially many value queries. Running
their algorithm on every prefix keeps this ratio and a polynomial number
of queries, but consecutive outputs can differ in up to $2k$ elements.
With polynomially many value queries, no algorithm exceeds
$0.470438681\ldots$ under a cardinality constraint, even offline
\citep[Lemma 8.5]{agglu26}. \citet[Lemma B.2]{buchbinder_chasing25} use exact
offline maximizers to maintain a $(1-\eps)$-approximate solution for
nonnegative subadditive functions under insertions with
$O(\eps^{-1}\log(k/\eps))$ amortized recourse. This bounds the total
number of changes over the stream, whereas we bound the number of changes
at each arrival.

\paragraph{Continuous relaxations and rounding.}
The multilinear extension and pipage rounding are standard tools in
submodular maximization \citep{calinescu11,chekuri10}. \citet{bian17}
relate approximately stationary points to global approximation and
obtain a $1/4$ guarantee for continuous submodular objectives with
diminishing returns over down-closed convex sets. We use approximate
stationarity to derive an inequality that later arrivals preserve
(Lemmas~\ref{lem:dual-certificate} and~\ref{lem:snapshot}). We round with
randomized pipage rounding, whose law does not depend on the objective,
so that elements can be added to a rounded set later
(Lemma~\ref{lem:rounding}).

\section{Model and preliminaries}\label{sec:model}

\subsection{Model}\label{sec:model-def}

Let $V=\{e_1,\ldots,e_n\}$ be a finite set of elements listed in arrival
order, and write $X_t=\{e_1,\ldots,e_t\}$ for the set of elements that
have arrived after $t$ arrivals, with $X_0=\varnothing$.
A function $f:2^V\to\mathbb R_{\ge0}$ is submodular if
\[
 f(A)+f(B)\ge f(A\cup B)+f(A\cap B)
 \qquad(A,B\subseteq V).
\]
The function is nonnegative on every subset of $V$, including sets with
more than $k$ elements, where $k\ge1$ is the cardinality budget. We do
not assume $f(\varnothing)=0$. For $e\notin A$, write
$f(e\mid A)=f(A\cup\{e\})-f(A)$. Submodularity is equivalent to
diminishing returns: $f(e\mid A)\ge f(e\mid B)$ whenever $A\subseteq B$
and $e\notin B$.

After the arrival of $e_t$, the algorithm reports a set $S_t$, and the
benchmark is the best feasible set of arrived elements:
\[
 S_t\subseteq X_t,\qquad |S_t|\le k,\qquad S_0=\varnothing,
 \qquad
 \OPT_t=\max_{\substack{Q\subseteq X_t\\|Q|\le k}}f(Q).
\]
The algorithm is an $\alpha$-approximation if $\E f(S_t)\ge\alpha\OPT_t$
for every time $t$, every function $f$, and every arrival order. Here $f$
and the order are fixed before the algorithm makes its random choices,
and the expectation is over those choices. The algorithm has recourse $C$
if $|S_t\symdiff S_{t-1}|\le C$ for every $t$ on every execution. The
algorithm may store all arrived elements and may reinsert elements that
it removed earlier. At time $t$, it may query $f(A)$ for any
$A\subseteq X_t$, including sets with more than $k$ elements. In the
full-foresight model, used for the upper bound, the algorithm knows the
arrival order and the value of $f$ on every subset of $V$ from the start,
but $S_t$ must still consist of arrived elements.

\subsection{Preliminaries}\label{sec:prelim}

For $z\in[0,1]^V$, let $R(z)$ be a random set that contains each element
$e$ independently with probability $z_e$. The multilinear extension of
$f$ is $F(z)=\E f(R(z))$. A vector on a subset $X\subseteq V$ is viewed
as a vector on $V$ with zero coordinates outside $X$. The function $F$
is linear in each coordinate. Hence its partial derivatives are
\begin{equation}\label{eq:first-derivative}
 \partial_eF(z)=\E\bigl[f\bigl(e\mid R(z)\setminus\{e\}\bigr)\bigr],
\end{equation}
which does not depend on $z_e$, and, for $e\ne e'$, with
$R=R(z)\setminus\{e,e'\}$,
\begin{equation}\label{eq:second-derivative}
 \partial_{e}\partial_{e'}F(z)
 =\E\bigl[f(R\cup\{e,e'\})-f(R\cup\{e\})-f(R\cup\{e'\})+f(R)\bigr]
 \le0,
\end{equation}
where the inequality is submodularity applied to $R\cup\{e\}$ and
$R\cup\{e'\}$. Linearity in each coordinate also gives
$\partial_e\partial_eF=0$. Two consequences of
\eqref{eq:second-derivative} are used repeatedly. First, $\partial_eF(z)$
is nonincreasing in every coordinate $z_{e'}$ with $e'\ne e$. Second, if
$d\in\mathbb R^V$ has nonnegative coordinates, then
$u\mapsto F(z+ud)$ is concave on every interval on which
$z+ud\in[0,1]^V$, because its second derivative is
$\sum_{e\ne e'}d_ed_{e'}\,\partial_e\partial_{e'}F(z+ud)\le0$.

For $y\in[0,1]^V$ and $Q\subseteq V$, write $y\vee\one_Q$ for the vector
that equals one on $Q$ and $y$ elsewhere. The following lemma compares a
fractional vector with small coordinates to an integral set.

\begin{lemma}[{\citealp[Lemma III.5]{feldman11}}]\label{lem:half-inclusion}
If $y\in[0,1/2]^V$, then $F(y\vee\one_Q)\ge\frac12 f(Q)$ for every
$Q\subseteq V$.
\end{lemma}

For $t\ge0$, let $M_t=\max\bigl(\{f(\varnothing)\}\cup\{f(\{e_j\}):j\le
t\}\bigr)$ be the largest value of a set of at most one arrived element.
Adding the elements of a nonempty arrived set $A$ one at a time and
applying diminishing returns gives
$f(A)\le f(\varnothing)+\sum_{e\in A}(f(\{e\})-f(\varnothing))
\le|A|M_t-(|A|-1)f(\varnothing)\le|A|M_t$. Together with
$f(\varnothing)\le M_t$, this gives
\begin{equation}\label{eq:arrived-value-bound}
 0\le f(A)\le\max\{|A|,1\}\,M_t
 \qquad(A\subseteq X_t).
\end{equation}
The empty set and every singleton are feasible because $k\ge1$, so
$M_t\le\OPT_t$.

The algorithm estimates derivatives by averaging samples. We use
Hoeffding's inequality in the following form. If $Z_1,\ldots,Z_m$ are
independent, take values in an interval of length $\ell$, and have mean
$\mu$, then
\begin{equation}\label{eq:hoeffding}
 \Pr\Bigl[\Bigl|\frac1m\sum_{i=1}^mZ_i-\mu\Bigr|>a\Bigr]
 \le2\exp\Bigl(-\frac{2ma^2}{\ell^2}\Bigr)
 \qquad(a>0).
\end{equation}

Finally, the algorithm turns fractional vectors into sets by randomized
pipage rounding. For the uniform matroid on a set $X$ whose independent
sets are the sets with at most $K$ elements, the matroid polytope is
$\{x\in[0,1]^X:\sum_{e\in X}x_e\le K\}$. Given $x$ in
this polytope, randomized pipage rounding returns a random set
$O\subseteq X$ with $|O|\le K$. Its distribution depends only on $x$,
and
\begin{equation}\label{eq:pipage}
 \E h(O)\ge H(x)
\end{equation}
for every submodular function $h:2^X\to\mathbb R$ with multilinear
extension $H$ \citep[Section~2 and Appendix~A]{chekuri10}. Elements are
added to a rounded set after it is drawn, so we need the following
consequence of \eqref{eq:pipage}.

\begin{lemma}[Rounding that permits later additions]\label{lem:rounding}
Let $X\subseteq V$, let $K\ge0$ be an integer, let $x\in[0,1]^X$ satisfy
$\sum_{e\in X}x_e\le K$, and let $O$ be the output of randomized pipage
rounding applied to $x$. Let $z\in[0,1]^{V\setminus X}$, and let
$Z\subseteq V\setminus X$ be a random set, independent of $O$, that
contains each element $e$ independently with probability $z_e$. Then
\begin{equation}\label{eq:rounding-future}
 \E f(O\cup Z)\ge F(x+z).
\end{equation}
\end{lemma}
\begin{proof}
Fix a set $Z_0\subseteq V\setminus X$, and define $h(A)=f(A\cup Z_0)$ for
$A\subseteq X$. For $A,A'\subseteq X$, submodularity of $f$ applied to
$A\cup Z_0$ and $A'\cup Z_0$ gives
$h(A)+h(A')\ge h(A\cup A')+h(A\cap A')$, so $h$ is submodular. Its
multilinear extension at $x$ is
$H(x)=\E f(R(x)\cup Z_0)=F(x+\one_{Z_0})$, where $R(x)\subseteq X$.
Hence \eqref{eq:pipage} gives
\begin{equation}\label{eq:fixed-addition}
 \E f(O\cup Z_0)\ge F(x+\one_{Z_0}).
\end{equation}
Since $Z$ is independent of $O$, conditioning on $Z=Z_0$ does not change
the distribution of $O$, and \eqref{eq:fixed-addition} gives
$\E[f(O\cup Z)\mid Z=Z_0]\ge F(x+\one_{Z_0})$. Taking the expectation
over $Z$ gives $\E f(O\cup Z)\ge\E F(x+\one_Z)$. The function $F$ is
linear in each coordinate of $V\setminus X$, and the coordinates of
$\one_Z$ are independent with means $z_e$, so $\E F(x+\one_Z)=F(x+z)$.
This proves \eqref{eq:rounding-future}.
\end{proof}

\section{Achievable result}\label{sec:algorithm}

\begin{theorem}[Approximation with constant recourse]
\label{thm:poly-quarter}
For every $\eps>0$, there is an algorithm that queries $f$ only on sets
of arrived elements and satisfies
\[
 \E f(S_t)\ge(1/4-\eps)\OPT_t
\]
at every time $t$, for every nonnegative submodular function $f$ and
every arrival order. Its recourse is $O(\eps^{-2})$ on every execution,
and it does not use the number of arrivals. For $0<\eps<1/4$, the
recourse is at most $8L^2+1$ with $L=\max\{2,\lceil1/\eps\rceil\}$, and
the number of value queries made up to time $N$ is
$O(N^8\eps^{-3}\log((N+1)/\eps))$.
\end{theorem}

The algorithm of Theorem~\ref{thm:poly-quarter} keeps a fractional
solution $y$, which assigns to every arrived element a selection
probability of at most one half. At random checkpoints it recomputes $y$
on all arrived elements, and between checkpoints it extends $y$ to each
new element without changing the old coordinates. Its candidate solution
consists of a rounded core of the checkpoint vector together with later
arrivals marked by independent coins; conditional on the numerical
computation, the expected value of the candidate solution is at least
$F(y)$ (Lemma~\ref{lem:candidate}). After each checkpoint, the reported
set moves from the previous candidate solution to the new one during a
window of $w$ arrivals.

\subsection{The algorithm}\label{sec:alg}

Fix $0<\eps<1/4$ and set
\begin{equation}\label{eq:schedule-parameters}
 \eta=\eps/8,\qquad L=\max\{2,\lceil1/\eps\rceil\}.
\end{equation}
If $k<4L^2$, the algorithm reports, after every arrival, the output of
Random Greedy \citep[Theorem 1.3]{bfns14} run with budget $k$ on
$X_t\cup D_t$, where $D_t$ is a set of $2k$ virtual dummy elements and
the objective is extended to $X_t\cup D_t$ by $f^+_t(A)=f(A\cap X_t)$;
the dummy elements are removed from the output. Their marginal values
are known to be zero, so every value query is on a subset of $X_t$. The rest of this section assumes $k\ge4L^2$ and uses the
integers
\begin{equation}\label{eq:block-parameters}
 w=\left\lfloor\frac{k}{2L^2}\right\rfloor,\qquad
 b=Lw,\qquad K=k-2b,\qquad
 r=\left\lceil\frac{2k}{w}\right\rceil.
\end{equation}
Here $w$ is the length of a replacement window, $b$ is the length of a
block of arrivals, $K$ is the budget of the checkpoint vector and of its
rounded core, and $r$ is the number of replacement operations allowed per
arrival. These parameters satisfy
\begin{equation}\label{eq:parameter-bounds}
 w\ge\frac{k}{4L^2},\qquad
 K\ge k\Bigl(1-\frac1L\Bigr)\ge\frac k2,\qquad
 r\le8L^2.
\end{equation}
The first bound holds because $k/(2L^2)\ge2$, and every real number
$c\ge2$ satisfies $\lfloor c\rfloor\ge c-1\ge c/2$. The second holds
because $2b=2Lw\le2L\cdot k/(2L^2)=k/L$ and $L\ge2$. The third holds
because the first bound gives $2k/w\le8L^2$, and $8L^2$ is an integer.
At time $t$, the estimates use the accuracy $\xi_t$, the failure
probability $\gamma_t$, and the sample size $m_t$ given by
\begin{equation}\label{eq:global-budgets}
 \xi_t=\frac{\eta M_t}{t(t+1)},\qquad
 \gamma_t=\frac{\eta}{4t(t+1)},\qquad
 m_t=\left\lceil
 \frac{t^2(t+1)^4}{2\eta^2}\log\frac{8t(t+1)}{\eta}
 \right\rceil,
\end{equation}
where $M_t$ is the scale defined in Section~\ref{sec:prelim}.

The algorithm uses four independent sources of randomness: one for the
checkpoint clocks, one for the numerical samples, one for rounding the
cores, and one for the marking coins. Ties in
Algorithm~\ref{alg:stationary} are broken by arrival order.

The algorithm recomputes the fractional solution at random checkpoints.
Block $j\ge0$ consists of the times $jb+1,\ldots,(j+1)b$. For each block
$j$, the algorithm draws $U_j$ uniformly from $\{0,\ldots,L-1\}$ and sets
\begin{equation}\label{eq:checkpoint-times}
 s_j=jb+U_jw,\qquad J_j=\{s_j+1,\ldots,s_j+w\}.
\end{equation}
Time $s_j$ is a checkpoint, and $J_j$ is its replacement window. Since
$b=Lw$, the $L$ possible windows
$\{jb+uw+1,\ldots,jb+(u+1)w\}$, $u=0,\ldots,L-1$, partition block $j$.
Because $|U_{j+1}-U_j|\le L-1$, consecutive checkpoints satisfy
\begin{equation}\label{eq:checkpoint-gaps}
 w\le s_{j+1}-s_j=b+(U_{j+1}-U_j)w\le2b-w.
\end{equation}
Time zero also serves as a checkpoint, with fractional solution $y=0$ and
an empty core and queue. For a time $t\ge1$, the active checkpoint is
the largest checkpoint $s$ with $s<t$.

At a checkpoint, the fractional solution is computed in a capped
polytope. For $X\subseteq V$ and an integer $K\ge1$, let
\begin{equation}\label{eq:capped-polytope}
 P_K(X)=\Bigl\{x\in[0,1/2]^X:\ \sum_{e\in X}x_e\le K\Bigr\}.
\end{equation}
At a checkpoint $s\ge1$, the algorithm calls Algorithm~\ref{alg:stationary}
(Section~\ref{sec:certificate}) on $X_s$ with budget $K$, accuracy
$\eta$, and failure probability $\gamma_s$. The call returns a
half-integral vector $x_s=\frac12\one_{A_s}\in P_K(X_s)$ and a number
$\lambda_s\ge0$; only the analysis uses $\lambda_s$. The algorithm sets
$y=x_s$ and draws a core $O_s\subseteq X_s$ with $|O_s|\le K$ by
randomized pipage rounding of $x_s$ (Section~\ref{sec:prelim}). For a
half-integral vector, with pairs taken in arrival order, this rounding
pairs the elements of $A_s$, keeps one element of each pair chosen by a
fair coin, and keeps a remaining unpaired element with probability one
half.

When $e_t$ arrives, its coordinate in $y$ is zero. The algorithm queries
$f(\{e_t\})$, updates $M_t$, and computes $\xi_t$, $\gamma_t$, and $m_t$
by \eqref{eq:global-budgets}. It estimates $g_t=\partial_{e_t}F(y)$ by
$\widehat g_t$, defined as zero if $M_t=0$ and otherwise as the average
of $m_t$ independent copies of $f(R\cup\{e_t\})-f(R)$, where
$R=R(y)\subseteq X_{t-1}$ is drawn afresh for each copy. It then sets
\begin{equation}\label{eq:fresh-rule}
 y_{e_t}=
 \begin{cases}
 1/2,&\widehat g_t>\xi_t,\\
 0,&\widehat g_t\le\xi_t,
 \end{cases}
\end{equation}
and leaves the other coordinates of $y$ unchanged. If $y_{e_t}=1/2$, the
algorithm flips a fair coin and, on heads, marks $e_t$ and inserts it
into the reported set.

The reported set moves to the new core gradually. At a checkpoint
$s\ge1$, after the output at time $s$ has been reported,
the algorithm freezes $B_s=S_s$ and forms a queue that first deletes the
elements of $B_s\setminus O_s$ and then inserts the elements of
$O_s\setminus B_s$. At each of the following arrivals, it performs the
next $r$ operations of the queue, or all remaining ones if fewer remain.
Algorithm~\ref{alg:main} lists the steps.

\begin{algorithm}[t]
\caption{Quarter approximation with constant recourse, for
$k\ge4L^2$}\label{alg:main}
\begin{algorithmic}[1]
\Require accuracy $\eps\in(0,1/4)$, budget $k$, value queries on
arrived elements
\State set $\eta,L$ by \eqref{eq:schedule-parameters} and $w,b,K,r$ by
\eqref{eq:block-parameters}
\State query $f(\varnothing)$ and set $M_0\gets f(\varnothing)$;
$y\gets0$, $S\gets\varnothing$, $\mathcal Q\gets$ empty queue
\State draw $U_0$ uniformly from $\{0,\ldots,L-1\}$ and set
$s_0\gets U_0w$
\For{$t=1,2,\ldots$}
 \If{$t=jb$ for some $j\ge1$}
  \State draw $U_j$ uniformly from $\{0,\ldots,L-1\}$ and set
  $s_j\gets jb+U_jw$
 \EndIf
 \State query $f(\{e_t\})$; set $M_t$, $\xi_t$, $\gamma_t$, $m_t$ by
 \eqref{eq:global-budgets}
 \State $\widehat g_t\gets0$ if $M_t=0$; otherwise $\widehat g_t\gets$
 average of $m_t$ samples of $f(R\cup\{e_t\})-f(R)$, $R\sim R(y)$
 \If{$\widehat g_t>\xi_t$}
  \State $y_{e_t}\gets1/2$
  \State with probability $1/2$: mark $e_t$ and set
  $S\gets S\cup\{e_t\}$\label{line:mark}
 \EndIf
 \State apply the next $\min\{r,|\mathcal Q|\}$ operations of
 $\mathcal Q$ to $S$\label{line:queue}
 \State report $S_t\gets S$
 \If{$t=s_j$ for some $j$}\label{line:checkpoint}
  \State $(x_t,\lambda_t)\gets$ Algorithm~\ref{alg:stationary} on $X_t$
  with budget $K$, accuracy $\eta$, failure probability $\gamma_t$
  \State $O_t\gets$ randomized pipage rounding of $x_t$
  \State $y\gets x_t$
  \State $\mathcal Q\gets$ deletions of $S\setminus O_t$, followed by
  insertions of $O_t\setminus S$
 \EndIf
\EndFor
\end{algorithmic}
\end{algorithm}

Line~\ref{line:checkpoint} runs after the output at time $t$ is
reported, so the output at a checkpoint still belongs to the previous
checkpoint. Since $s_0\le(L-1)w<b$ and $s_j\ge jb$, every checkpoint
time is known before it occurs.

The analysis considers one checkpoint interval at a time. Fix a
checkpoint $s$, and let $s'$ be the next checkpoint. Within this
interval, $y_s=x_s$, with $x_0=0$, and for $s<t\le s'$, $y_t$ denotes the
fractional solution after the arrival at time $t$ has been processed. At
the shared endpoint $s'$, the vector $y_{s'}$ of this interval is the
vector before the new computation; the new vector is $x_{s'}$. The
estimates satisfy $g_t=\partial_{e_t}F(y_{t-1})$, where $y_{t-1}$ has a
zero coordinate at $e_t$. The candidate solution at a time $t$ with
$s\le t\le s'$ is
\[
 C_t=O_s\cup A_t,
\]
where $O_0=\varnothing$ and $A_t$ is the set of elements marked at times
$s+1,\ldots,t$.

\subsection{The value of the fractional solution}\label{sec:certificate}

The analysis of a checkpoint uses a linear program over $P_K(X)$. For
$x\in P_K(X)$, write $g=\nabla F(x)\in\mathbb R^X$ for the vector of
partial derivatives in the coordinates of $X$. The vector $x$ is a
stationary point of $F$ on $P_K(X)$ if
$\max_{z\in P_K(X)}\inner{g}{z-x}=0$. For a number $\lambda\ge0$, define
\begin{align}
 U_K(g,\lambda)&=K\lambda+\frac12\sum_{e\in X}(g_e-\lambda)_+,
 \label{eq:dual-objective}\\
 \Gamma_K(x,\lambda)&=U_K(g,\lambda)-\inner{x}{g},
 \label{eq:dual-gap}
\end{align}
where $c_+=\max\{c,0\}$. The linear program
$\max\{\inner{g}{z}:z\in P_K(X)\}$ has the dual program
$\min\{U_K(g,\lambda):\lambda\ge0\}$, so $\Gamma_K(x,\lambda)$ is the
gap between a dual value and the linearized value $\inner{x}{g}$. If $x$
is a stationary point and $\lambda$ is an optimal dual solution, this
gap is zero; Lemma~\ref{lem:sampled-stationarity} computes a pair with a
small gap.

For a time $t$, a vector $y\in[0,1]^{X_t}$, and numbers $\ell,\lambda\ge0$,
define
\begin{equation}\label{eq:D-definition}
 D_\ell(y;\lambda)
 =\ell\lambda+
 \sum_{e\in X_t}
 \bigl[(1-y_e)(\partial_eF(y))_+-\lambda\bigr]_+.
\end{equation}
The quantity $(1-y_e)(\partial_eF(y))_+$ is an upper bound on the gain of
raising $y_e$ to one, and $D_\ell(y;\lambda)$ bounds the total of these
gains over any $\ell$ elements. The next lemma bounds $D_K$ at a
checkpoint by the value of the fractional solution and the gap.

\begin{lemma}[The inequality at a checkpoint]\label{lem:dual-certificate}
Let $X=X_s$ for a time $s$. For every $x\in P_K(X)$ and $\lambda\ge0$,
with $\Gamma=\Gamma_K(x,\lambda)$,
\begin{equation}\label{eq:initial-certificate}
 \Gamma\ge0,\qquad
 D_K(x;\lambda)\le F(x)+2\Gamma,\qquad
 K\lambda\le F(x)+\Gamma.
\end{equation}
\end{lemma}
\begin{proof}
Write $g=\nabla F(x)$. We first rewrite $\Gamma$. Substituting
$x_eg_e=x_e\lambda+x_e(g_e-\lambda)$ into \eqref{eq:dual-gap} gives
\[
 \Gamma=\lambda\Bigl(K-\sum_{e\in X}x_e\Bigr)
 +\sum_{e\in X}\Bigl[\tfrac12(g_e-\lambda)_+-x_e(g_e-\lambda)\Bigr].
\]
The bracket equals $(1/2-x_e)(g_e-\lambda)$ when $g_e>\lambda$ and equals
$x_e(\lambda-g_e)$ when $g_e\le\lambda$. Hence
\begin{equation}\label{eq:gap-decomposition}
 \Gamma=\lambda\Bigl(K-\sum_{e\in X}x_e\Bigr)
 +\sum_{e:\,g_e>\lambda}(1/2-x_e)(g_e-\lambda)
 +\sum_{e:\,g_e\le\lambda}x_e(\lambda-g_e).
\end{equation}
Since $x\in P_K(X)$ and $\lambda\ge0$, each of the three terms in
\eqref{eq:gap-decomposition} is nonnegative. This proves $\Gamma\ge0$.
It also shows that the middle term of \eqref{eq:gap-decomposition} is at
most $\Gamma$.

Next we bound each summand of $D_K(x;\lambda)$ in \eqref{eq:D-definition}.
We claim that, for every $e\in X$,
\begin{equation}\label{eq:summand-bound}
 \bigl[(1-x_e)(g_e)_+-\lambda\bigr]_+\le(1-x_e)(g_e-\lambda)_+.
\end{equation}
If $g_e\le\lambda$, then $(g_e)_+\le\lambda$ because $\lambda\ge0$, so
$(1-x_e)(g_e)_+-\lambda\le0$ and the left side of
\eqref{eq:summand-bound} is zero. If $g_e>\lambda$, then $g_e>0$ and
\[
 (1-x_e)g_e-\lambda=(1-x_e)(g_e-\lambda)-x_e\lambda
 \le(1-x_e)(g_e-\lambda),
\]
and the right side is nonnegative, so \eqref{eq:summand-bound} holds
again. Summing \eqref{eq:summand-bound} over $e\in X$, adding $K\lambda$,
and writing $1-x_e=1/2+(1/2-x_e)$ gives
\[
 D_K(x;\lambda)
 \le K\lambda+\frac12\sum_{e\in X}(g_e-\lambda)_+
 +\sum_{e:\,g_e>\lambda}(1/2-x_e)(g_e-\lambda)
 \le U_K(g,\lambda)+\Gamma,
\]
where the last step uses \eqref{eq:dual-objective} and the bound on the
middle term of \eqref{eq:gap-decomposition}. By \eqref{eq:dual-gap},
$U_K(g,\lambda)=\inner{x}{g}+\Gamma$, so
\begin{equation}\label{eq:D-via-gradient}
 D_K(x;\lambda)\le\inner{x}{g}+2\Gamma.
\end{equation}

It remains to show $\inner{x}{g}\le F(x)$. The function
$\varphi(u)=F(ux)$ is concave on $[0,1]$, because $x$ has nonnegative
coordinates (Section~\ref{sec:prelim}). Also
$\varphi(0)=f(\varnothing)\ge0$ and $\varphi'(1)=\inner{x}{g}$. For a
concave function, the derivative at the right endpoint is at most the
slope of the chord, so
$\inner{x}{g}=\varphi'(1)\le\varphi(1)-\varphi(0)\le F(x)$. Substituting
into \eqref{eq:D-via-gradient} proves the second inequality of
\eqref{eq:initial-certificate}. For the third, the definition
\eqref{eq:dual-objective} gives $K\lambda\le U_K(g,\lambda)$, and
therefore $K\lambda\le\inner{x}{g}+\Gamma\le F(x)+\Gamma$.
\end{proof}

Algorithm~\ref{alg:stationary} makes the gap in
Lemma~\ref{lem:dual-certificate} small, using value queries on arrived
elements only. It keeps a half-integral vector $x=\frac12\one_A$. In
each iteration it estimates every partial derivative of $F$ at $x$ by
sampling, takes the $2K$ elements with the largest positive estimates,
and compares the estimated linearized value of this choice with that of
$x$. If the difference is small, it returns $x$. Otherwise it takes one
of the half-integral steps from $x$ toward this choice (adding one
element, removing one, or exchanging two), the one with the largest
estimated gain. It uses the parameters
\begin{equation}\label{eq:stationarity-parameters}
 a=\frac{\eta M}{8\nu},\qquad
 T=\left\lceil\frac{2\nu^2}{\eta}\right\rceil+1,\qquad
 m=\left\lceil
       \frac{32\nu^2(\nu+1)^2}{\eta^2}
       \log\frac{2\nu T}{\gamma}
    \right\rceil,
\end{equation}
where $\nu=|X|$, $a$ is the accuracy of each estimated derivative, $T$ is
the largest number of iterations, and $m$ is the number of samples per
derivative. The bound in Lemma~\ref{lem:sampled-stationarity} is
relative to the scale $M=M_s$, which the algorithm observes.

\begin{algorithm}[t]
\caption{Approximately stationary fractional solution}
\label{alg:stationary}
\begin{algorithmic}[1]
\Require set $X$ with $\nu=|X|$, integer budget $K\ge1$, accuracy
$\eta\in(0,1)$, failure probability $\gamma\in(0,1)$
\State $M\gets\max\bigl(\{f(\varnothing)\}\cup\{f(\{e\}):e\in X\}\bigr)$
\If{$\nu=0$ or $M=0$} \Return $(x,\lambda)=(0,0)$ \EndIf
\State set $a,T,m$ by \eqref{eq:stationarity-parameters};
$A\gets\varnothing$ \Comment{the current vector is $x=\frac12\one_A$}
\For{$i=1,\ldots,T$}
 \For{each $e\in X$}
  \State $\widehat g_e\gets$ average of $m$ samples of
  $f(R\cup\{e\})-f(R)$, with $R=R(x)\setminus\{e\}$ drawn afresh
 \EndFor
 \State $Y\gets$ the $2K$ largest positive coordinates of $\widehat g$,
 or all positive ones if fewer; $z\gets\frac12\one_Y$
 \State $\lambda\gets$ the $(2K+1)$-th largest coordinate of
 $\widehat g$ if it exists and is positive, and $\lambda\gets0$ otherwise
 \State $\widehat\Gamma\gets\inner{\widehat g}{z-x}$
 \If{$\widehat\Gamma\le3\eta M/4$ or $i=T$} \Return $(x,\lambda)$
 \EndIf\label{line:stop}
 \State pair the elements of $Y\setminus A$ with those of
 $A\setminus Y$ in arrival order; each pair $(b,c)$ gives the move
 $d=\frac12(\one_b-\one_c)$, each unpaired $b\in Y\setminus A$ gives
 $d=\frac12\one_b$, and each unpaired $c\in A\setminus Y$ gives
 $d=-\frac12\one_c$\label{line:moves}
 \State apply to $x$ a move $d$ with the largest
 $\inner{\widehat g}{d}$, and update $A$ accordingly
\EndFor
\end{algorithmic}
\end{algorithm}

\begin{lemma}[Computing the fractional solution]
\label{lem:sampled-stationarity}
Let $X=X_s$ with $\nu=|X|$, and let $M=M_s$. Let $0<\eta,\gamma<1$ and
let $K\ge1$ be an integer. On every execution,
Algorithm~\ref{alg:stationary} returns a vector $x=\frac12\one_A$ with
$A\subseteq X$ and $|A|\le2K$, so that $x\in P_K(X)$, together with a
number $\lambda\ge0$, and it makes
\[
 O\!\left(\nu^7\eta^{-3}\log\frac{\nu+1}{\eta\gamma}\right)
\]
value queries, all on subsets of $X$. Conditional on the history before
the call, $\Gamma_K(x,\lambda)\le\eta M$ with probability at least
$1-\gamma$.
\end{lemma}
The proof is provided in Appendix~\ref{app:stationarity}.

The next lemma shows that the inequality of
Lemma~\ref{lem:dual-certificate} survives the arrivals that follow a
checkpoint. In the ideal case $K=k$, with zero gap and exact estimates,
the lemma gives $D_k(y;\lambda)\le F(y)$ at every later time. Together
with Lemma~\ref{lem:half-inclusion} and the bound
$F(y\vee\one_Q)\le F(y)+D_k(y;\lambda)$, which its proof establishes,
this gives, for an optimal set $Q$,
\[
 \frac12\OPT\le F(y\vee\one_Q)\le F(y)+D_k(y;\lambda)\le2F(y),
\]
that is, $F(y)\ge\OPT/4$. The probability one half is what keeps the
inequality valid after an arrival: it raises $F$ by $g/2$ for a new
element with derivative $g>0$, which pays for the new term
$[g/2-\lambda]_+$ of $D_k$. Started from $x=0$ and $\lambda=0$, the same
argument gives $F(y)\ge\OPT/4$, but a set drawn from $y$ can then have
more than $k$ elements; this is close to the $1/4$ guarantee of a
uniformly random set without a constraint \citep{feige11}. The
checkpoints and the budget $K$ make the argument compatible with the
cardinality constraint and with bounded recourse. The reduced
budget $K<k$, the gap, and the estimation errors are the losses that the
lemma keeps track of; the replacement windows add the factor $1-1/L$ in
Section~\ref{sec:theorem-proof}. The bound holds even if the optimum
grows by a large factor after the checkpoint, which matters for
non-monotone objectives.

\begin{lemma}[Approximation between checkpoints]\label{lem:snapshot}
Let $s$ and $s'$ be consecutive checkpoints, and let $1\le K\le k$.
Suppose that the numbers $\lambda\ge0$ and $\Gamma_s\ge0$ satisfy
$\Gamma_K(x_s,\lambda)\le\Gamma_s$, and that
$|\widehat g_\tau-g_\tau|\le\xi_\tau$ for $s<\tau\le s'$. Then, for
$s<t\le s'$,
\begin{equation}\label{eq:snapshot-ratio}
 F(y_t)\ge
 \frac{K}{2(k+K)}\OPT_t
 -\Gamma_s-\frac{2K}{k+K}\sum_{\tau=s+1}^t\xi_\tau.
\end{equation}
\end{lemma}
\begin{proof}
For $s\le\tau\le s'$, let $\Phi_\tau=D_K(y_\tau;\lambda)-F(y_\tau)$, where
$D_K$ is taken over $X_\tau$ in \eqref{eq:D-definition}. We first show
that, for $s<\tau\le s'$,
\begin{equation}\label{eq:potential-step}
 \Phi_\tau\le\Phi_{\tau-1}+2\xi_\tau,
 \qquad F(y_\tau)\ge F(y_{\tau-1}).
\end{equation}
The vectors $y_\tau$ and $y_{\tau-1}$ differ only in the coordinate of
$e_\tau$, which is zero in $y_{\tau-1}$. The sum defining
$D_K(y_\tau;\lambda)$ has one more term than the sum defining
$D_K(y_{\tau-1};\lambda)$, namely the term of $e_\tau$. By
\eqref{eq:first-derivative}, $\partial_{e_\tau}F(y_\tau)=g_\tau$.

Suppose first that $\widehat g_\tau>\xi_\tau$, so that
$(y_\tau)_{e_\tau}=1/2$. Since $|\widehat g_\tau-g_\tau|\le\xi_\tau$, we
have $g_\tau>0$. Linearity of $F$ in the coordinate of $e_\tau$ gives
$F(y_\tau)-F(y_{\tau-1})=g_\tau/2>0$. For an old element $e\in
X_{\tau-1}$, raising the coordinate of $e_\tau$ does not increase
$\partial_eF$ (Section~\ref{sec:prelim}), so the term of $e$ in $D_K$
does not increase. The new term equals
$[(1-1/2)g_\tau-\lambda]_+=[g_\tau/2-\lambda]_+\le g_\tau/2$. Hence $D_K$
increases by at most $g_\tau/2$, while $F$ increases by exactly
$g_\tau/2$, and $\Phi_\tau\le\Phi_{\tau-1}$.

Suppose next that $\widehat g_\tau\le\xi_\tau$, so that
$y_\tau=y_{\tau-1}$ on $X_{\tau-1}$ and $(y_\tau)_{e_\tau}=0$. Then
$F(y_\tau)=F(y_{\tau-1})$, and every old term of $D_K$ is unchanged. The
accuracy assumption gives $g_\tau\le\widehat g_\tau+\xi_\tau\le2\xi_\tau$,
so the new term is $[(g_\tau)_+-\lambda]_+\le(g_\tau)_+\le2\xi_\tau$.
Hence $\Phi_\tau\le\Phi_{\tau-1}+2\xi_\tau$. The two cases prove
\eqref{eq:potential-step}.

Summing \eqref{eq:potential-step} over $\tau=s+1,\ldots,t$ and applying
Lemma~\ref{lem:dual-certificate} to $x_s$, which gives
$\Phi_s\le2\Gamma_K(x_s,\lambda)\le2\Gamma_s$, yields
\begin{equation}\label{eq:D-evolution}
 D_K(y_t;\lambda)
 \le F(y_t)+2\Gamma_s+2\sum_{\tau=s+1}^t\xi_\tau.
\end{equation}
The third inequality of \eqref{eq:initial-certificate} and the
monotonicity in \eqref{eq:potential-step} give
\begin{equation}\label{eq:lambda-bound}
 K\lambda\le F(x_s)+\Gamma_s\le F(y_t)+\Gamma_s.
\end{equation}
By \eqref{eq:D-definition}, $D_k(y_t;\lambda)=D_K(y_t;\lambda)+(k-K)\lambda$.
Bounding the first term by \eqref{eq:D-evolution} and the second by
$(k-K)/K$ times \eqref{eq:lambda-bound} gives
\begin{equation}\label{eq:D-k-bound}
 D_k(y_t;\lambda)
 \le\frac{k}{K}F(y_t)
     +\frac{k+K}{K}\Gamma_s
     +2\sum_{\tau=s+1}^t\xi_\tau.
\end{equation}

Now fix $Q\subseteq X_t$ with $|Q|\le k$ and $f(Q)=\OPT_t$, and list
its elements as $v_1,\ldots,v_p$. Let $z^0=y_t$, and let $z^i$ be $z^{i-1}$
with the coordinate of $v_i$ raised to one, so that $z^p=y_t\vee\one_Q$.
The coordinate of $v_i$ in $z^{i-1}$ is $(y_t)_{v_i}$, so linearity in
this coordinate gives
$F(z^i)-F(z^{i-1})=(1-(y_t)_{v_i})\,\partial_{v_i}F(z^{i-1})$. The vector
$z^{i-1}$ is at least $y_t$ in every coordinate other than $v_i$, so
$\partial_{v_i}F(z^{i-1})\le\partial_{v_i}F(y_t)\le(\partial_{v_i}F(y_t))_+$.
Summing over $i$ gives
\begin{equation}\label{eq:raise-Q}
 F(y_t\vee\one_Q)
 \le F(y_t)+\sum_{e\in Q}c_e,
 \qquad c_e=(1-(y_t)_e)(\partial_eF(y_t))_+.
\end{equation}
Each $c_e$ satisfies $c_e\le\lambda+(c_e-\lambda)_+$. Summing over $Q$,
using $|Q|\le k$, and extending the sum of positive parts to all of $X_t$
gives $\sum_{e\in Q}c_e\le D_k(y_t;\lambda)$.

All coordinates of $y_t$ are at most one half, because $x_s\in P_K(X_s)$
and \eqref{eq:fresh-rule} sets new coordinates to $0$ or $1/2$.
Lemma~\ref{lem:half-inclusion}, \eqref{eq:raise-Q}, and
\eqref{eq:D-k-bound} therefore give
\[
 \frac12\OPT_t\le F(y_t\vee\one_Q)
 \le\frac{k+K}{K}F(y_t)+\frac{k+K}{K}\Gamma_s
 +2\sum_{\tau=s+1}^t\xi_\tau.
\]
Multiplying by $K/(k+K)$ and rearranging proves \eqref{eq:snapshot-ratio}.
\end{proof}

The candidate solution inherits the value of the fractional solution.
To state this, let $\mathcal N$ be the $\sigma$-field generated by the
checkpoint clocks and the numerical random source. Each fractional
solution $y_t$ is determined by $\mathcal N$, while the cores and the
marking coins use the other two independent sources.

\begin{lemma}[Value of the candidate solution]\label{lem:candidate}
Let $s$ be a checkpoint with next checkpoint $s'$, and let
$s<t\le s'$. Then $\E[f(C_t)\mid\mathcal N]\ge F(y_t)$.
\end{lemma}
\begin{proof}
Condition on $\mathcal N$. Then $x_s$ and $y_t$ are fixed. The vector
$y_t$ equals $x_s$ on $X_s$, and on $X_t\setminus X_s$ it equals a
vector $z$ whose coordinates are $1/2$ for the elements with
$\widehat g_\tau>\xi_\tau$ and $0$ for the others. Thus $y_t=x_s+z$.
The core $O_s$ is drawn by pipage rounding of $x_s$ using the rounding
source, so its conditional law is the law in Lemma~\ref{lem:rounding}.
The set $A_t$ contains each element of $X_t\setminus X_s$ with
$z_e=1/2$ independently with probability $1/2$, and it is independent of
$O_s$. For $s=0$ the core is empty and $x_0=0$, and the same description
holds. Applying \eqref{eq:rounding-future} with $X=X_s$, $x=x_s$, and
$Z=A_t$ gives $\E[f(O_s\cup A_t)\mid\mathcal N]\ge F(x_s+z)=F(y_t)$.
\end{proof}

\subsection{Feasibility, recourse, and accuracy}\label{sec:transport}

The size and recourse bounds depend only on the counting of queue
operations and marked elements. They hold on every execution, including
executions on which the estimates are inaccurate.

\begin{lemma}[Feasibility and recourse]\label{lem:transport}
Assume $k\ge4L^2$. On every execution of Algorithm~\ref{alg:main},
$|S_t|\le k$ and $|S_t\symdiff S_{t-1}|\le r+1\le8L^2+1$ for every $t$.
Moreover, if the active checkpoint $s$ of a time $t$ satisfies $s=0$ or
$t\ge s+w$, then $S_t=C_t$.
\end{lemma}
\begin{proof}
Let $s$ and $s'$ be consecutive checkpoints, where $s=0$ is allowed. We
prove by induction over the checkpoints that, for every $t$ with
$s<t\le s'$, the reported set has at most $k$ elements, and that it
equals $C_t$ if $s=0$ or $t\ge s+w$. The induction hypothesis at $s\ge1$
is that $S_s=C_s$, where $C_s$ is the candidate solution of the
checkpoint before $s$.

For $s=0$ there is no queue, so $S_t=A_t=C_t$ for $0<t\le s'$. If
$s_0\ge1$, then $s'=s_0\le(L-1)w=b-w$. If $s_0=0$, then $s'=s_1\le2b-w$
by \eqref{eq:checkpoint-gaps}. In both cases $|S_t|\le t\le2b-w\le k$.

Now let $s\ge1$, and let $s_-$ be the previous checkpoint. By the
induction hypothesis, $B_s=S_s=C_s=O_{s_-}\cup A_s$. The core $O_{s_-}$
has at most $K$ elements, and $A_s$ contains only elements that arrived
at times $s_-+1,\ldots,s$. There are at most $2b-w$ such times, by
\eqref{eq:checkpoint-gaps} if $s_-\ge1$ and by the bound $s\le2b-w$ of
the previous paragraph if $s_-=0$. Hence
\begin{equation}\label{eq:frozen-size}
 |B_s|\le K+2b-w=k-w.
\end{equation}
The queue has $|B_s\setminus O_s|+|O_s\setminus B_s|\le|B_s|+|O_s|\le2k$
operations, and the algorithm performs $r$ of them per arrival, with
$rw\ge2k$ by \eqref{eq:block-parameters}. So the queue is exhausted by
the end of the update at time $s+w$, and $s+w\le s'$ by
\eqref{eq:checkpoint-gaps}.

Next we describe the reported set at a time $t$ with $s<t\le s'$. The
elements that arrive after $s$ are not in $B_s\cup O_s$, and the queue
operates only on $B_s\cup O_s$. Hence $S_t=P_t\cup A_t$, where
$P_t\subseteq X_s$ is the part produced by the queue. While deletions
remain in the queue, $P_t\subseteq B_s$. After the last deletion,
including within the update that performs it, $P_t\subseteq O_s$. Once
the queue is empty, $P_t=O_s$.

We now bound the size. For $t\le s+w$, at most $w$ elements have been
marked since $s$, so $|A_t|\le w$. While deletions remain,
\eqref{eq:frozen-size} gives $|S_t|\le|B_s|+w\le k$. After the last
deletion, $|S_t|\le|O_s|+w\le K+w\le k$, since $K+w=k-2b+w$. For
$t\ge s+w$, the queue is empty after the update at time $t$, so
$S_t=O_s\cup A_t=C_t$; moreover $|A_t|\le t-s\le s'-s\le2b-w$ by
\eqref{eq:checkpoint-gaps}, so $|S_t|\le K+2b=k$. In particular, since
$s'\ge s+w$, we have $S_{s'}=C_{s'}$, which is the induction hypothesis
at $s'$.

The recourse bound follows because each update performs at most $r$
queue operations and inserts at most one marked element. A queue is
formed only after the output at its checkpoint is reported, and the
previous queue is empty by then, so no update performs operations of
two queues. Hence $|S_t\symdiff S_{t-1}|\le r+1\le8L^2+1$ by
\eqref{eq:parameter-bounds}. The last claim of the lemma is the second
part of the induction statement.
\end{proof}

The estimates are accurate with high probability. The parameters in
\eqref{eq:global-budgets} are indexed by the global time, so a single
union bound covers all checkpoints and arrivals, and the algorithm does
not need the number of arrivals.

\begin{lemma}[Accuracy of all estimates]\label{lem:global-accuracy}
Let $\mathcal G$ be the event that every call of
Algorithm~\ref{alg:stationary} made at a checkpoint $s\ge1$ returns a
pair with $\Gamma_K(x_s,\lambda_s)\le\eta M_s$, and that every estimate
made at a time $t$ satisfies $|\widehat g_t-g_t|\le\xi_t$. Conditional on
any realization of the checkpoint clocks, $\mathcal G$ has probability
at least $1-\eta/2$. On $\mathcal G$, for every checkpoint $s$ with next
checkpoint $s'$ and every $s<t\le s'$,
\begin{equation}\label{eq:successful-snapshot}
 F(y_t)\ge
 \Bigl(\frac{K}{2(k+K)}-2\eta\Bigr)\OPT_t.
\end{equation}
Up to time $N$, Algorithm~\ref{alg:main} makes
$O(N^8\eta^{-3}\log((N+1)/\eta))$ value queries.
\end{lemma}
\begin{proof}
We first bound the failure probability of one estimate $\widehat g_t$.
If $M_t=0$, then \eqref{eq:arrived-value-bound} gives $f(A)=0$ for every
$A\subseteq X_t$, so $g_t=0=\widehat g_t$. Otherwise fix the clocks and
the numerical history before time $t$. Then $y_{t-1}$ is fixed, and the
fresh numerical randomness used at time $t$ produces $m_t$ independent
samples from the required product distribution. By
\eqref{eq:first-derivative}, each sample $f(R\cup\{e_t\})-f(R)$ has mean
$g_t$. By submodularity and $f(\varnothing)\ge0$, each sample is at most
$f(\{e_t\})-f(\varnothing)\le M_t$, and by \eqref{eq:arrived-value-bound}
it is at least $-f(R)\ge-tM_t$. The samples therefore lie in an interval
of length at most $(t+1)M_t$. With $a=\xi_t$, \eqref{eq:hoeffding} and
\eqref{eq:global-budgets} give
\[
 \Pr\bigl[|\widehat g_t-g_t|>\xi_t\bigm|\text{clocks, history}\bigr]
 \le2\exp\Bigl(-\frac{2m_t\eta^2}{t^2(t+1)^4}\Bigr)
 \le2\cdot\frac{\eta}{8t(t+1)}=\gamma_t,
\]
since $2m_t\eta^2/(t^2(t+1)^4)\ge\log(8t(t+1)/\eta)$.

At a checkpoint $s\ge1$, the call uses fresh numerical randomness, so
Lemma~\ref{lem:sampled-stationarity} with failure probability $\gamma_s$
shows that $\Gamma_K(x_s,\lambda_s)\le\eta M_s$ fails with probability at
most $\gamma_s$, conditional on the clocks and the earlier numerical
history. At each time there is at most one estimate and at most one call
of the routine of Lemma~\ref{lem:sampled-stationarity}. A union bound
and the identity $1/(t(t+1))=1/t-1/(t+1)$ give
\[
 \Pr[\text{not }\mathcal G\mid\text{clocks}]
 \le\sum_{t\ge1}2\gamma_t
 =\frac\eta2\sum_{t\ge1}\Bigl(\frac1t-\frac1{t+1}\Bigr)
 =\frac\eta2.
\]

Now assume $\mathcal G$. For the checkpoint $s=0$, the vector $x_0=0$
lives on $X_0=\varnothing$, so $\Gamma_K(x_0,0)=0$ by
\eqref{eq:dual-gap}; set $\lambda_0=0$ and $\Gamma_0=0$. For $s\ge1$, set
$\Gamma_s=\eta M_s$. Since $M_\tau$ is nondecreasing in $\tau$, the same
identity gives
\[
 \sum_{\tau=s+1}^t\xi_\tau
 \le\eta M_t\sum_{\tau=s+1}^t\Bigl(\frac1\tau-\frac1{\tau+1}\Bigr)
 \le\eta M_t.
\]
Lemma~\ref{lem:snapshot} with $\lambda=\lambda_s$ applies, and together
with $\Gamma_s\le\eta M_s\le\eta M_t$, $2K/(k+K)\le1$, and
$M_t\le\OPT_t$ it gives
\[
 F(y_t)\ge\frac{K}{2(k+K)}\OPT_t-\eta M_t-\eta M_t
 \ge\Bigl(\frac{K}{2(k+K)}-2\eta\Bigr)\OPT_t.
\]

Finally we count queries up to time $N$. The call at a checkpoint
$t\le N$ makes $O(t^7\eta^{-3}\log((t+1)/(\eta\gamma_t)))$ queries by
Lemma~\ref{lem:sampled-stationarity}. Since $1/\gamma_t=4t(t+1)/\eta$,
the logarithm is $O(\log((N+1)/\eta))$. Summing over at most $N$
checkpoints gives $O(N^8\eta^{-3}\log((N+1)/\eta))$. The estimates use
$2m_t$ queries at time $t$, which sum to
$O(N^7\eta^{-2}\log((N+1)/\eta))$, and the singleton queries and the
query of $f(\varnothing)$ add $N+1$. Both terms are smaller than the
first. The cores are drawn without queries.
\end{proof}

\subsection{Proof of Theorem~\ref{thm:poly-quarter}}
\label{sec:theorem-proof}

\begin{proof}
If $\eps\ge1/4$, the empty output satisfies the theorem. Let
$0<\eps<1/4$, and choose $\eta$ and $L$ by \eqref{eq:schedule-parameters}.

Suppose first that $k<4L^2$, and use the Random Greedy branch of
Section~\ref{sec:alg}. With $2k$ dummy elements available, Random Greedy
returns a set of at most $k$ elements of $X_t\cup D_t$ whose expected
value is at least $\OPT_t/e$ for the extended objective, and the
extended objective agrees with $f$ after the dummy elements are removed
\citep[Reduction 1 and Theorem 1.3]{bfns14}. It makes $O(tk)$ value
queries, all on subsets of $X_t$. Since $1/e>1/4$, the ratio holds. Two
sets of at most $k$ elements differ in at most $2k<8L^2$ elements, which
bounds the recourse. The total number of queries up to time $N$ is
$O(N^2k)=O(N^2\eps^{-2})$.

Suppose now that $k\ge4L^2$, and run Algorithm~\ref{alg:main}.
Lemma~\ref{lem:transport} gives $|S_t|\le k$ and recourse at most
$8L^2+1=O(\eps^{-2})$ on every execution, and
Lemma~\ref{lem:global-accuracy} with $\eta=\eps/8$ gives the stated
number of queries. Every query is on a subset of the arrived elements,
and no step uses the number of arrivals.

It remains to prove the ratio at a fixed time $t$. Let $j$ be the block
containing $t$, condition on all clocks, and suppose that $t\notin J_j$.
Let $s$ be the active checkpoint of $t$. If $s=0$, then
Lemma~\ref{lem:transport} gives $S_t=C_t$. Otherwise, either $s$ is the
checkpoint of an earlier block, whose window lies in that block, so that
$t>s+w$; or $s=s_j$, and then $t>s_j+w$ because $t\notin J_j$. In both
cases Lemma~\ref{lem:transport} gives $S_t=C_t$.

Fix clocks with $t\notin J_j$. Condition further on $\mathcal N$. By
Lemma~\ref{lem:candidate}, $\E[f(S_t)\mid\mathcal N]\ge F(y_t)$. On the
event $\mathcal G$ of Lemma~\ref{lem:global-accuracy}, the bound
\eqref{eq:successful-snapshot} applies; off $\mathcal G$, $F(y_t)\ge0$.
The coefficient in \eqref{eq:successful-snapshot} is positive, because
$K\ge k/2$ gives $K/(2(k+K))\ge1/6$ while $2\eta<1/16$. Taking the
expectation over $\mathcal N$ given the clocks, and using
$\Pr[\mathcal G\mid\text{clocks}]\ge1-\eta/2\ge1-\eta$, gives
\[
 \E[f(S_t)\mid\text{clocks}]
 \ge(1-\eta)\Bigl(\frac{K}{2(k+K)}-2\eta\Bigr)\OPT_t.
\]

The event $t\notin J_j$ depends only on $U_j$. Since the $L$ possible
windows partition block $j$, it has probability $1-1/L$. If $t\in J_j$,
then $f(S_t)\ge0$. Averaging over the clocks therefore gives
\begin{equation}\label{eq:finite-ratio}
 \E f(S_t)\ge
 \Bigl(1-\frac1L\Bigr)(1-\eta)
 \Bigl(\frac{K}{2(k+K)}-2\eta\Bigr)\OPT_t.
\end{equation}

We now bound the coefficient in \eqref{eq:finite-ratio}. The function
$K\mapsto K/(2(k+K))$ is increasing, so \eqref{eq:parameter-bounds}
gives
\[
 \frac{K}{2(k+K)}\ge\frac{1-1/L}{2(2-1/L)}\ge\frac{1-1/L}{4}
 =\frac14-\frac1{4L},
\]
where the second step uses $2-1/L\le2$. Let
$c=K/(2(k+K))-2\eta$. Then $0<c\le1/4$, because $K\le k$. Also
$(1-1/L)(1-\eta)\ge1-1/L-\eta$. Hence the coefficient in
\eqref{eq:finite-ratio} is at least
\[
 \Bigl(1-\frac1L-\eta\Bigr)c
 \ge c-\frac14\Bigl(\frac1L+\eta\Bigr)
 \ge\frac14-\frac1{4L}-2\eta-\frac1{4L}-\frac\eta4
 =\frac14-\frac1{2L}-\frac94\eta.
\]
By \eqref{eq:schedule-parameters}, $1/(2L)\le\eps/2$ and
$9\eta/4=9\eps/32$, so the coefficient is at least
$1/4-25\eps/32\ge1/4-\eps$.
\end{proof}

\section{Upper bound result}\label{sec:upper}

The hard instance is the cut function of a weighted directed graph whose
elements arrive in groups of decreasing sizes. It adapts the instance of
\citet[Appendix A]{buchbinder19}, in which each arrival is a single
vertex. The graph and the order are fixed in advance, so the bound holds
in the full-foresight model. Write $H_m=\sum_{i=1}^m1/i$ and
$H_m^{(2)}=\sum_{i=1}^m1/i^2$.

\begin{theorem}[A quarter upper bound with full foresight]
\label{thm:quarter-upper}
Let $C\ge0$, let $q,\rho,m\ge2$ be integers, and let $k\ge q^{m-1}$. There
is a directed-cut instance with budget $k$ such that every algorithm in
the full-foresight model, with unrestricted queries and computation,
that has recourse at most $C$ on every execution and satisfies
$\E f(S_t)\ge\alpha\OPT_t$ at every time $t$ of this instance has
\begin{equation}\label{eq:quarter-upper-finite}
 \alpha\le\frac14+\frac{C}{q-1}+\frac1{\rho-1}
 +\frac{H_m^{(2)}}{2(H_m-1)}.
\end{equation}
Consequently, if an algorithm in this model is an $\alpha$-approximation
on every instance and its recourse is at most a constant $C$,
independent of $k$ and $n$, then $\alpha\le1/4$.
\end{theorem}

The proof uses two ratios. The ratio $q$ between consecutive group sizes
bounds how much the reported fraction of an old group can change after
the group arrives, which gives the error $C/(q-1)$. The ratio $\rho$
between the total weights of the arcs leaving consecutive groups bounds
the contribution of the older groups when a new group completes, which
gives the error $1/(\rho-1)$. The reported fractions then satisfy a
recursion whose squared form telescopes, and the last term in
\eqref{eq:quarter-upper-finite}, which tends to zero as $m$ grows, forces
the coefficient $1/4$.

\begin{proof}
We first define the instance. The ground set consists of disjoint groups
$G_1,\ldots,G_m$ of sizes
\[
 d_i=q^{m-i},
\]
so that every group has at most $d_1=q^{m-1}\le k$ elements. The
elements arrive group by group: first $G_1$, then $G_2$, and so on, in
any fixed order within each group. Let $t_i=\sum_{l=1}^id_l$ be the time
at which group $i$ completes. Set $W_1=0$ and $W_i=\rho^{\,i-2}$ for
$i\ge2$. For every $i\ge2$, every $j<i$, every $a\in G_i$, and every
$v\in G_j$, the graph has an arc from $a$ to $v$ of weight
\[
 \frac{W_i}{(i-1)d_id_j}.
\]
There are no other arcs. Let $f(S)$ be the total weight of the arcs whose
tail is in $S$ and whose head is not in $S$. A cut function of a directed
graph with nonnegative weights is nonnegative and submodular, and
$f(\varnothing)=0$. The graph and the order do not depend on the
algorithm.

The arcs leaving $G_i$ have total weight
$d_i\sum_{j<i}d_j\cdot W_i/((i-1)d_id_j)=W_i$, and their heads lie in
earlier groups, outside $G_i$. Hence $f(G_i)=W_i$. Since $|G_i|\le k$ and
$G_i$ has arrived by time $t_i$,
\begin{equation}\label{eq:quarter-group-benchmark}
 \OPT_{t_i}\ge W_i\qquad(i\ge2).
\end{equation}

Next we express the value of the reported set. For an execution of the
algorithm and $j\le i$, define the fraction of group $j$ in the reported
set at time $t_i$, and the fraction at the completion of group $j$, by
\[
 \theta_{j,i}=\frac{|S_{t_i}\cap G_j|}{d_j},\qquad
 \theta_j=\theta_{j,j}.
\]
All these numbers lie in $[0,1]$. At time $t_i$, the reported set
contains only elements of $G_1,\ldots,G_i$, and the elements of $G_1$
have no outgoing arcs. Fix $2\le l\le i$. Each element of
$S_{t_i}\cap G_l$ is the tail of $d_j(1-\theta_{j,i})$ arcs into
$G_j\setminus S_{t_i}$ for each $j<l$, each of weight
$W_l/((l-1)d_ld_j)$. Summing over the $d_l\theta_{l,i}$ such elements and
over $j<l$ gives
\begin{equation}\label{eq:cut-identity}
 f(S_{t_i})=
 \sum_{l=2}^{i} W_l\,\theta_{l,i}
 \Bigl(1-\frac{1}{l-1}\sum_{j<l}\theta_{j,i}\Bigr).
\end{equation}

The fraction of an old group changes little. Let $j<i$. Between times
$t_j$ and $t_i$ there are $t_i-t_j$ arrivals, each with at most $C$
changes, so $\bigl||S_{t_i}\cap G_j|-|S_{t_j}\cap G_j|\bigr|\le
C(t_i-t_j)$. Since
$t_i-t_j=\sum_{l=j+1}^id_l\le\sum_{l=j+1}^m q^{m-l}\le d_j/(q-1)$,
\begin{equation}\label{eq:quarter-density-drift}
 |\theta_{j,i}-\theta_j|
 \le\frac{C(t_i-t_j)}{d_j}
 \le\frac{C}{q-1}.
\end{equation}

We now bound the value in terms of the completion fractions. For
$i=2,\ldots,m+1$, let
\[
 \phi_i=1-\frac{1}{i-1}\sum_{j<i}\theta_j\in[0,1].
\]
We bound the two parts of \eqref{eq:cut-identity} at time $t_i$. The
term $l=i$ equals
$W_i\theta_i(1-\frac1{i-1}\sum_{j<i}\theta_{j,i})$. By
\eqref{eq:quarter-density-drift}, the average of $\theta_{j,i}$ over
$j<i$ is at least the average of $\theta_j$ minus $C/(q-1)$, so this term
is at most $W_i\theta_i(\phi_i+C/(q-1))\le W_i(\theta_i\phi_i+C/(q-1))$.
Each term $l<i$ is at most $W_l$, because the fractions lie in $[0,1]$.
Their sum satisfies
\[
 \sum_{l=2}^{i-1}W_l=\sum_{l=2}^{i-1}\rho^{\,l-2}
 =\frac{\rho^{\,i-2}-1}{\rho-1}\le\frac{W_i}{\rho-1},
\]
where the sum is empty for $i=2$. Therefore, on every execution, with
$\Delta=C/(q-1)+1/(\rho-1)$,
\begin{equation}\label{eq:quarter-output-bound}
 f(S_{t_i})\le W_i(\theta_i\phi_i+\Delta)\qquad(i\ge2).
\end{equation}
The approximation guarantee at time $t_i$, together with
\eqref{eq:quarter-group-benchmark} and \eqref{eq:quarter-output-bound},
gives $\alpha W_i\le\E f(S_{t_i})\le W_i(\E[\theta_i\phi_i]+\Delta)$.
Dividing by $W_i>0$ gives
\begin{equation}\label{eq:quarter-product-bound}
 \E[\theta_i\phi_i]\ge\alpha-\Delta
 \qquad(i=2,\ldots,m).
\end{equation}
The fractions $\theta_1,\ldots,\theta_m$ may depend on each other in any
way.

We now show that the products $\theta_i\phi_i$ cannot all be large, on
every execution. By the definition of $\phi_i$,
$(i-1)(1-\phi_i)=\sum_{j<i}\theta_j$ and
$i(1-\phi_{i+1})=\sum_{j\le i}\theta_j$. Subtracting gives
$i(1-\phi_{i+1})=(i-1)(1-\phi_i)+\theta_i$, which rearranges to
\[
 \phi_{i+1}=\phi_i+\frac{c_i}{i},\qquad c_i=1-\theta_i-\phi_i.
\]
Squaring this identity gives
\[
 \phi_{i+1}^2-\phi_i^2=\frac{2\phi_ic_i}{i}+\frac{c_i^2}{i^2}
 =\frac{2}{i}\bigl(\phi_i(1-\phi_i)-\theta_i\phi_i\bigr)
 +\frac{c_i^2}{i^2}.
\]
Since $\phi_i\in[0,1]$, we have $\phi_i(1-\phi_i)\le1/4$. Since
$\theta_i,\phi_i\in[0,1]$, we have $|c_i|\le1$. Hence
\[
 \phi_{i+1}^2-\phi_i^2
 \le\frac{2}{i}\Bigl(\frac14-\theta_i\phi_i\Bigr)+\frac1{i^2}.
\]
Summing over $i=2,\ldots,m$, the left side telescopes to
$\phi_{m+1}^2-\phi_2^2\ge-1$. Rearranging and using
$1+\sum_{i=2}^m i^{-2}=H_m^{(2)}$ gives
\[
 2\sum_{i=2}^{m}\frac{\theta_i\phi_i-1/4}{i}\le H_m^{(2)}.
\]

Taking expectations, applying \eqref{eq:quarter-product-bound}, and using
$\sum_{i=2}^m i^{-1}=H_m-1$ gives
\[
 2\Bigl(\alpha-\Delta-\frac14\Bigr)(H_m-1)\le H_m^{(2)}.
\]
Dividing by $2(H_m-1)>0$ proves \eqref{eq:quarter-upper-finite}.

For the second claim, let $C$ be a constant independent of $k$ and $n$.
For each $m\ge2$ and all integers $q,\rho\ge2$, the instance above with
$k=q^{m-1}$ gives \eqref{eq:quarter-upper-finite} with the same $C$.
Letting $q$ and $\rho$ tend to infinity with $m$ fixed gives
$\alpha\le1/4+H_m^{(2)}/(2(H_m-1))$. As $m\to\infty$, $H_m^{(2)}$ stays
below $\pi^2/6$ while $H_m$ diverges, so $\alpha\le1/4$. Along these
instances $C$ stays fixed, while $k$, $n$, and the weights $W_i$ grow.
\end{proof}

The same construction shows that recourse must grow as a power of the
budget to exceed $1/4$ by a fixed amount.

\begin{corollary}[Recourse needed above one quarter]
\label{cor:poly-recourse}
Let $0<\beta\le1$, let $m_\beta=\lceil\exp(1+\pi^2/(3\beta))\rceil$, and
let $c_\beta=1/(m_\beta-1)$. Suppose that, for a budget
$k\ge2^{m_\beta-1}$, an algorithm in the full-foresight model satisfies
$\E f(S_t)\ge(1/4+\beta)\OPT_t$ at every time of every instance with
budget $k$ and has recourse at most $C_k$ on every such execution. Then
\[
 C_k\ge\frac\beta2\bigl(\lfloor k^{c_\beta}\rfloor-1\bigr).
\]
In particular, recourse $k^{o(1)}$ does not allow any fixed ratio above
$1/4$.
\end{corollary}
\begin{proof}
Apply Theorem~\ref{thm:quarter-upper} with $C=C_k$, $m=m_\beta$,
$q=\lfloor k^{c_\beta}\rfloor$, and $\rho=1+\lceil4/\beta\rceil$. Since
$k^{c_\beta}\ge2$, we have $q\ge2$, and $q^{m-1}\le k$. The choice of
$\rho$ gives $1/(\rho-1)\le\beta/4$. Since $H_m^{(2)}\le\pi^2/6$ and
$H_m\ge\ln(m+1)>\ln m\ge1+\pi^2/(3\beta)$, we have
$H_m-1\ge\pi^2/(3\beta)$, and therefore
\[
 \frac{H_m^{(2)}}{2(H_m-1)}\le\frac{\pi^2}6\cdot\frac{3\beta}{2\pi^2}
 =\frac\beta4.
\]
Substituting these bounds and $\alpha=1/4+\beta$ into
\eqref{eq:quarter-upper-finite} gives
$1/4+\beta\le1/4+C_k/(q-1)+\beta/2$, that is, $C_k\ge(\beta/2)(q-1)$.
\end{proof}

The proof of Theorem~\ref{thm:quarter-upper} uses the budget only through
$|G_i|\le k$ in \eqref{eq:quarter-group-benchmark}, and it never uses a
bound on $|S_t|$. With $k=n$, every set is feasible, so the bound holds
even when the cardinality constraint does not bind.

\section{Conclusion}\label{sec:conclusion}
With constant recourse, the optimal limiting ratio for non-monotone
submodular maximization under a cardinality constraint is $1/4$.
Polynomially many queries on arrived elements achieve $1/4-\eps$ with
recourse $O(\eps^{-2})$, and no algorithm exceeds $1/4$ even with full
foresight and unlimited computation; any fixed ratio above $1/4$ requires
recourse that grows as a positive power of $k$. Two questions remain open: the smallest
recourse that achieves $1/4-\eps$ as a function of $\eps$, and the
smallest number of queries.

\acks{GPT (OpenAI) was used to review the proofs, and Claude (Anthropic) was
used to revise the text of the manuscript. The author checked all content
and takes full responsibility for the article.}

\appendix
\section{Proof of Lemma~\ref{lem:sampled-stationarity}: computing an
approximately stationary fractional solution}
\label{app:stationarity}

\begin{proof}
In Algorithm~\ref{alg:stationary}, the vector $z$ maximizes
$\inner{\widehat g}{z}$ over $P_K(X)$, and $\lambda$ minimizes
$U_K(\widehat g,\lambda)$ over $\lambda\ge0$. These two linear programs
are dual to each other, so their optimal values are equal. With
\eqref{eq:dual-gap}, this gives
\begin{equation}\label{eq:estimated-gap}
 \widehat\Gamma=\inner{\widehat g}{z}-\inner{\widehat g}{x}
 =U_K(\widehat g,\lambda)-\inner{x}{\widehat g}.
\end{equation}

We first treat the early return. If $\nu=0$, all sums over $X$ are empty
and $\Gamma_K(0,0)=0$. If $M=0$, then \eqref{eq:arrived-value-bound}
gives $f(A')=0$ for every $A'\subseteq X$, so $\nabla F(0)=0$ and
$\Gamma_K(0,0)=U_K(0,0)=0$. In both cases the lemma holds, with at most
$\nu+1$ queries. From now on, $\nu\ge1$ and $M>0$.

The vector $x$ stays of the form $\frac12\one_A$ with $|A|\le2K$ on
every execution. This holds initially. A move from a pair keeps $|A|$
unchanged, and a move $-\frac12\one_c$ decreases it. A move
$\frac12\one_b$ occurs only if $Y\setminus A$ has more elements than
$A\setminus Y$, that is, if $|Y|>|A|$; then $|A|\le|Y|-1\le2K-1$, and the
move keeps $|A|\le2K$. Hence every vector is in $P_K(X)$, and the
returned $\lambda$ is nonnegative by construction.

Next we count queries. There are at most $T$ iterations, each with one
test at line~\ref{line:stop} and $\nu$ estimates, $m$ samples per
estimate, and two queries per sample, in addition to $\nu+1$ queries for
$M$. Since $T\le2\nu^2/\eta+2\le4\nu^2/\eta$ and
$\log(2\nu T/\gamma)\le\log(8\nu^3/(\eta\gamma))
=O(\log((\nu+1)/(\eta\gamma)))$, \eqref{eq:stationarity-parameters} gives
\[
 2\nu Tm+\nu+1=O\Bigl(\nu\cdot\frac{\nu^2}{\eta}\cdot\frac{\nu^4}{\eta^2}
 \log\frac{\nu+1}{\eta\gamma}\Bigr)
 =O\Bigl(\nu^7\eta^{-3}\log\frac{\nu+1}{\eta\gamma}\Bigr).
\]
This bound holds on every execution, and every queried set is a subset
of $X$.

We now bound the probability that some estimate is inaccurate. Write
$g=\nabla F(x)$ for the current $x$. Conditional on the history before an
estimate, the samples of $\widehat g_e$ are independent, and by
\eqref{eq:first-derivative} each has mean $g_e$. By submodularity and
$f(\varnothing)\ge0$, each sample is at most
$f(\{e\})-f(\varnothing)\le M$, and by \eqref{eq:arrived-value-bound} it
is at least $-f(R)\ge-\nu M$. The samples therefore lie in an interval of
length at most $(\nu+1)M$. By \eqref{eq:hoeffding} and
\eqref{eq:stationarity-parameters},
\[
 \Pr\bigl[|\widehat g_e-g_e|>a\bigm|\text{history}\bigr]
 \le2\exp\Bigl(-\frac{2ma^2}{(\nu+1)^2M^2}\Bigr)
 =2\exp\Bigl(-\frac{m\eta^2}{32\nu^2(\nu+1)^2}\Bigr)
 \le\frac{\gamma}{\nu T}.
\]
There are at most $\nu T$ estimates, so a union bound shows that the
event $\mathcal E$ that every estimate satisfies $|\widehat g_e-g_e|\le
a$ has probability at least $1-\gamma$, conditional on the history before
the call. For the rest of the proof assume $\mathcal E$.

At every test, the computed $\widehat\Gamma$ is close to the true gap of
the current pair $(x,\lambda)$. By \eqref{eq:dual-gap} and
\eqref{eq:estimated-gap},
\[
 \Gamma_K(x,\lambda)-\widehat\Gamma
 =\bigl(U_K(g,\lambda)-U_K(\widehat g,\lambda)\bigr)
 -\inner{x}{g-\widehat g}.
\]
The map $c\mapsto(c-\lambda)_+$ is $1$-Lipschitz, so the first
difference is at most $\frac12\sum_e|g_e-\widehat g_e|$ in absolute
value. Since $0\le x_e\le1/2$, the second term is also at most
$\frac12\sum_e|g_e-\widehat g_e|$ in absolute value. Hence
\begin{equation}\label{eq:gap-error}
 |\Gamma_K(x,\lambda)-\widehat\Gamma|\le\sum_{e\in X}|g_e-\widehat g_e|
 \le\nu a=\frac{\eta M}8.
\end{equation}
If the test at line~\ref{line:stop} returns because
$\widehat\Gamma\le3\eta M/4$, then \eqref{eq:gap-error} gives
$\Gamma_K(x,\lambda)\le7\eta M/8<\eta M$, as required.

It remains to show that, on $\mathcal E$, such a return happens before
the test with $i=T$. Consider a test that does not return, so
$\widehat\Gamma>3\eta M/4$. The moves formed at line~\ref{line:moves}
sum to $z-x=\frac12\one_{Y\setminus A}-\frac12\one_{A\setminus Y}$, and
there are at most $\nu$ of them. Hence the chosen move $d$ satisfies
$\inner{\widehat g}{d}\ge\widehat\Gamma/\nu>3\eta M/(4\nu)$. Every move
has $\sum_e|d_e|\le1$, so
\[
 \inner{g}{d}\ge\inner{\widehat g}{d}-\sum_{e\in X}|g_e-\widehat g_e|\,|d_e|
 \ge\inner{\widehat g}{d}-a
 >\frac{3\eta M}{4\nu}-\frac{\eta M}{8\nu}
 >\frac{\eta M}{2\nu}.
\]
We next show that $F(x+d)-F(x)\ge\inner{g}{d}$. If $d$ changes a single
coordinate, then $F$ is linear along $d$ and equality holds. If
$d=\frac12(\one_b-\one_c)$, let $\psi(u)=F(x+ud)$ for $u\in[0,1]$; the
points $x+ud$ lie in $[0,1]^X$. Since $\partial_b\partial_bF=
\partial_c\partial_cF=0$, the second derivative is
$\psi''(u)=2d_bd_c\,\partial_b\partial_cF(x+ud)
=-\frac12\partial_b\partial_cF(x+ud)\ge0$ by
\eqref{eq:second-derivative}. So $\psi$ is convex, and
$\psi(1)-\psi(0)\ge\psi'(0)=\inner{g}{d}$. In both cases, every test that
does not return raises $F$ by more than $\eta M/(2\nu)$.

Suppose the tests $i=1,\ldots,T-1$ all failed to return. Starting from
$F(0)=f(\varnothing)\ge0$, the $T-1$ moves would give
$F(x)>(T-1)\eta M/(2\nu)\ge\nu M$, using $T-1\ge2\nu^2/\eta$. But $F(x)$
is an average of values $f(A')$ with $A'\subseteq X$, so $F(x)\le\nu M$
by \eqref{eq:arrived-value-bound}. This contradiction shows that, on
$\mathcal E$, some test with $i\le T-1$ returns with
$\widehat\Gamma\le3\eta M/4$, and the returned pair satisfies
$\Gamma_K(x,\lambda)\le\eta M$.
\end{proof}

\bibliography{references}
\end{document}